\documentclass[twocolumn]{aa}

\usepackage[varg]{txfonts}
\usepackage{amsmath}
\usepackage{graphicx}
\usepackage{comment}
\usepackage{color} 
\usepackage{natbib}
\newcommand{\fig}[1]{Fig.~\ref{#1}}

\newcommand{\sect}[1]{Sect.~\ref{#1}}
\newcommand{\pone}{Paper I}
\newcommand{\ptwo}{Paper II}

\begin{document}
\title{Solar magnetic flux rope eruptions caused by inverse flux feeding processes}
\author{
Quanhao Zhang\inst{\ref{inst1},\ref{inst2},\ref{inst3}}\and
Shangbin Yang\inst{\ref{inst4},\ref{inst5}}\and
Rui Liu\inst{\ref{inst1},\ref{inst2},\ref{inst3}}\and
Min Zhang\inst{\ref{inst1},\ref{inst6}}\and
Dong Wang\inst{\ref{inst6}}\and
Ake Zhao\inst{\ref{inst7}}\and
Shaoyu Lyu\inst{\ref{inst1},\ref{inst2}}\and
Anchuan Song\inst{\ref{inst1},\ref{inst2},\ref{inst8}}\and
Yuming Wang\inst{\ref{inst1},\ref{inst2},\ref{inst8},\ref{inst9}} 
}
\institute{School of Earth and Space Sciences, University of Science and Technology of China, Hefei 230026, China\\ \email{zhangqh@ustc.edu.cn}\label{inst1}
\and
CAS Center for Excellence in Comparative Planetology/CAS Key Laboratory of Geospace Environment/Mengcheng National Geophysical Observatory, University of Science and Technology of China, Hefei 230026, China\label{inst2}
\and
Collaborative Innovation Center of Astronautical Science and Technology, Hefei 230026, China\label{inst3}
\and
Key Laboratory of Solar Activity, National Astronomical Observatories, Chinese Academy of Sciences, 100012 Beijing, China\label{inst4}
\and
University of Chinese Academy of Sciences, 100049 Beijing, China\label{inst5}
\and
Anhui Jianzhu University, Hefei 230026, China\label{inst6}
\and
College of Physics and Electric Information, Luoyang Normal University, Luoyang, Henan 471934, China\label{inst7}
\and
National Key Laboratory of Deep Space Exploration, University of Science and Technology of China, Hefei 230026, China\label{inst8}
\and
Hefei National Laboratory, University of Science and Technology of China, Hefei 230088, China\label{inst9}
}
\abstract{Large-scale solar eruptions are generally accepted to have coronal magnetic flux ropes as their core structures. Recent studies found that the solar eruptions could be initiated by a sequence of flux feeding processes, during with chromospheric fibrils rise and merge with the pre-existing coronal flux rope. Further theoretical analyses have demonstrated that the normal flux feeding, i.e. the axial magnetic flux within the fibril is in the same direction as that in the flux rope, results in the accumulation of the total axial flux within the flux rope, so as to initiate the eruption. If the directions of the axial flux in the fibril and the flux rope are opposite, it is termed inverse flux feeding, whose influence on coronal flux ropes, however, is still unclear. In this paper, we use a 2.5-dimensional magnetohydrodynamic model to simulate the evolution of coronal flux ropes associated with inverse flux feeding. It is found that inverse flux feeding is also efficient in causing solar eruptions: although the total signed axial magnetic flux of the rope decreases after inverse flux feeding, the total unsigned axial flux can accumulate; the eruption occurs if the unsigned axial flux of the rope reaches a critical value, which is almost the same as the threshold for normal flux feeding. The total axial currents within the rope are also similar during the onset of the eruptions caused by both normal and inverse flux feeding. Our simulation results suggest that it is the unsigned axial magnetic flux rather than the signed axial flux that regulates the onset of coronal flux rope eruptions.}
\titlerunning{Solar eruptions caused by inverse flux feeding}
\maketitle

\section{Introduction}
\label{sec:introduction}
With the development of science and technology, the impact of space weather on our human beings is becoming increasingly obvious \citep{Schwenn2006a,Temmer2021,Su2021,Su2023}. It is generally accepted that large-scale solar eruptive activities are the primary source of extreme space weather \citep{svestka2001a,Cheng2014,Patsourakos2020,Jiang2023}. The radiation, energetic particles, and ejected magnetized plasma produced by solar eruptions have profound effects on not only the solar-terrestrial but also the planetary space environment \citep{Guo2018,Green2018,Li2022,Ye2023} . Large-scale solar eruptive activities include filament/prominence eruptions \citep{Li2016,Jenkins2018,Fan2020,Li2025}, flares \citep{Shibata2011a,Li2016a,Zhang2019}, and coronal mass ejections \citep[CMEs,][]{Lugaz2017,Veronig2018,Owens2020,Mei2023}. They are not independent of each other, but are usually considered as specific manifestations of coronal flux rope eruptions at different spatial and temporal scales \citep{Zhang2001,Lin2003a,Vrvsnak2005a,Guo2010a,Chen2020a}. Typically, after a coronal flux rope is initiated to erupt, the prominence/filament constrained inside the rope also rises along with the rope, resulting in the eruption of the prominence/filament; a current sheet is formed beneath the flux rope during the eruption, so that magnetic energy is drastically released and converted to thermal energy and particle acceleration via magnetic reconnection in the current sheet, which is observed as a flare; the erupted flux rope propagates in the coronal and interplanetary space, and its observed counterpart is a CME. Therefore, coronal magnetic flux ropes are the core structures of the large-scale solar eruptive activities \citep{Gopalswamy2018a,Liu2020a,Chen2023}. The investigation on the driving mechanism and evolutionary scenario of the coronal flux rope eruptions is of great scientific significance for not only understanding solar eruptions, but also ensuring space weather safety.
\par
To shed light on the triggering mechanism of coronal flux rope eruptions, many theoretical models were proposed in previous studies. In these models, the onset of the eruption are correlated with various observational phenomena, such as photospheric flux emergence \citep{Toriumi2014,Syntelis2019,Li2023}, collisional shear \citep{Chintzoglou2019,Toeroek2024}, sunspot rotation \citep{Bi2016,Vemareddy2016,Yan2018}, and flux feeding \citep{Zhang2014,Zhang2020}. These processes gradually accumulate energy within the flux rope system, ultimately leading to a loss of equilibrium or instability in the system, which results in the onset of the eruption. The corresponding physical mechanism dominating the onset also differs, which could be ideal magnetohydrodynamic (MHD) instabilities \citep{Torok2003a,Kliem2006a,Savcheva2012b,Keppens2019,Ledentsov2021}, magnetic reconnection \citep{Antiochos1999a,Chen2000a,Moore2001a,Archontis2008b,Inoue2018}, or flux rope catastrophes \citep{vanTend1978a,Forbes1995a,Demoulin2010a,Longcope2014a,Zhang2016a,Zhang2021}.  
\par
\begin{figure*}
	\includegraphics[width=\hsize]{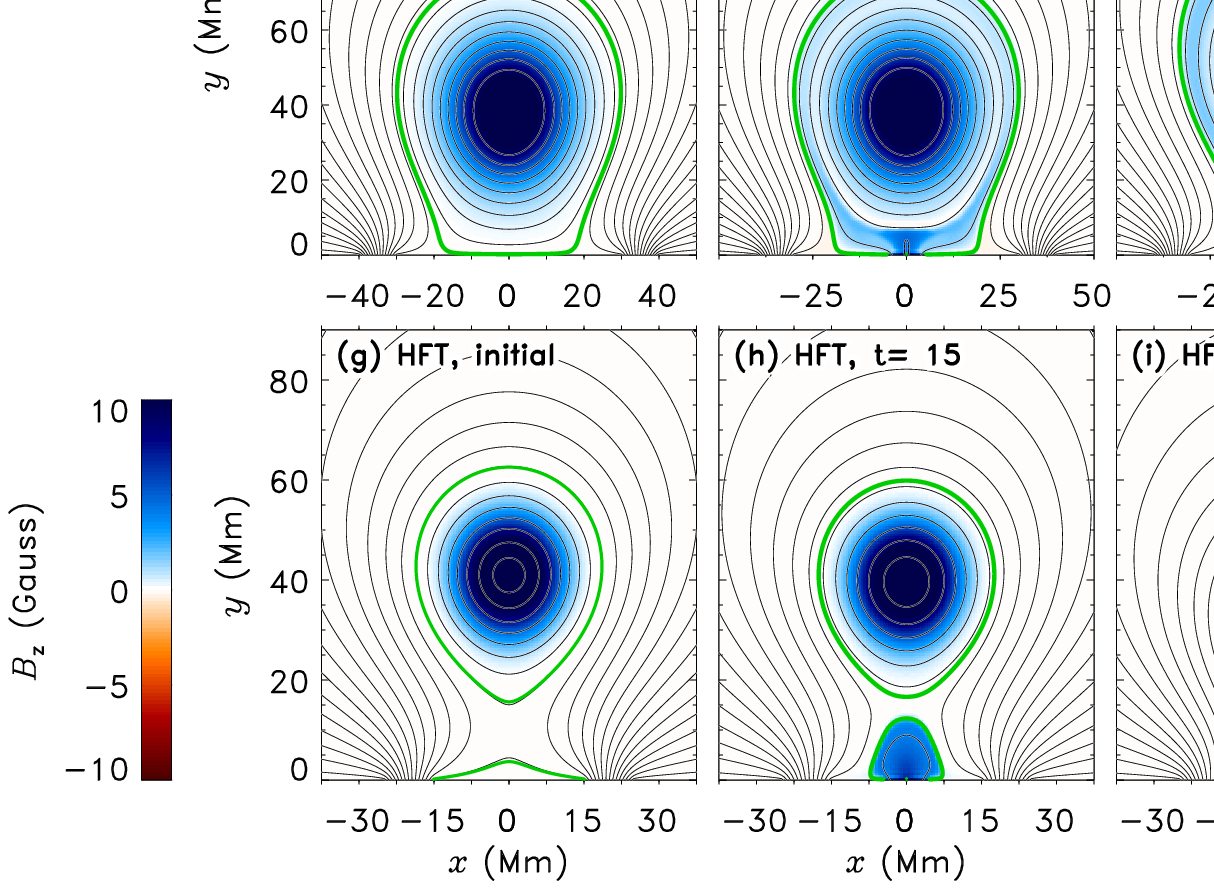}
	\caption{Coronal flux rope eruptions caused by normal flux feeding. Top panels are the simulation results of flux ropes in the BPS configuration, which are adapted from \cite{Zhang2020}; bottom panels are the results for the HFT configuration, adapted from \cite{Zhang2021a}.The black curves in panels (a)-(e) and (g)-(k) illustrate the temporal evolution of the magnetic field configuration; the green curves mark the boundary of the flux rope, and that of the emerging fibril (panel (h)); the blue-red color depicts the distribution of the axial magnetic flux in $z-$direction. Panel (f) and panel (l) are the evolutions of the height of the flux rope axis.  The vertical dotted lines in panel (f) and panel (g) correspond to the times of panels (a)-(e) and panels (g)-(k), respectively. }\label{fig:old}
\end{figure*}

It is observed by \cite{Zhang2014} that solar eruptions could be caused by flux feeding processes, during which chromospheric fibrils rise and merge with a solar prominence, activating the prominence and eventually causing the eruption. Numerical simulations are further carried out to investigate the physical scenario of the eruption caused by flux feeding. The simulation results are illustrated in \fig{fig:old}, which are adapted from \cite{Zhang2020} (hereafter \pone, top panels) and \cite{Zhang2021a} (hereafter \ptwo, bottom panels), respectively.
The black curves plot the magnetic field lines, and the blue-red color depicts the distribution of $B_z$, i.e., the axial component of the magnetic field. Here \fig{fig:old}(a) is the initial state for the simulation in \pone: the flux rope sticks to the photosphere, wrapped by a bald patch separatrix surface \citep[BPSS,][]{Titov1993a,Gibson2006a}. This is one of the two typical types of coronal flux ropes, known as the BPS configuration. Obviously, there is only positive (blue color) axial magnetic flux distributed around the center of the ropes in the initial BPS configuration. As shown in \fig{fig:old}(b)-\ref{fig:old}(c), a small flux rope representing the rising fibril emerges from the photosphere into the pre-existing coronal flux rope; the axial magnetic flux within the fibril is also positive, so that positive axial flux is injected into the rope from its lower boundary during flux feeding, and then distributed within the outer section of the flux rope. It is demonstrated by \pone ~that flux feeding results in the accumulation of the axial magnetic flux within the flux rope, and if its total axial flux exceeds the critical value $\Phi_{zc}^B$ of the order $1.2\times10^{20}$ Mx, the rope is initiated to erupt, as shown in \fig{fig:old}(d)-\ref{fig:old}(e) and \fig{fig:old}(f). The bottom panels in \fig{fig:old} are the simulation results in \ptwo, and \fig{fig:old}(g) is the corresponding initial state. This is the other type of coronal flux rope system, in which the rope is suspended in the corona, with coronal arcades and X point below the rope, so that is called the HFT (Hyperbolic Flux Tube) configuration \citep{Titov2003,Aulanier2005,Chintzoglou2017}. As demonstrated by \ptwo, flux feeding also injects positive axial flux into the pre-existing flux rope in the HFT configuration (\fig{fig:old}(h)-\ref{fig:old}(i)), and the rope is initiated to erupt if its total axial flux exceeds the critical value $\Phi_{zc}^H$ of the order $7.1\times10^{19}$ Mx (\fig{fig:old}(j)-\ref{fig:old}(k) and \fig{fig:old}(l)). Based on the simulation results introduced above, it is concluded by \pone ~and \ptwo ~that flux feeding is efficient in causing coronal flux rope eruptions.
\par
It is noteworthy that in all the previous simulations about flux feeding, the direction of the axial magnetic field within the rising fibril is the same as that within the pre-existing flux rope, so that the total axial magnetic flux of the rope always accumulates after this kind of flux feeding. However, many observational studies demonstrate that the chirality and helicity of newly emerging magnetic flux could be quite different from that of the magnetic system within the surrounding active region \citep[e.g.,][]{Zhang2001a,Yang2009,Cheung2014,vanDriel2015a}. This indicate that the axial magnetic field within chromospheric fibrils in actual solar corona might not always be the same as that within the pre-existing flux rope. In other words, flux feeding could also inject axial flux into the flux rope in the opposite direction, which is thus referred to as ``inverse'' flux feeding in the rest of this paper. For comparison, the flux feeding processes investigated in previous simulations are termed as ``normal'' flux feeding. Different from normal flux feeding, the total axial flux of the rope might not always accumulate after inverse flux feeding. It is still unclear how the flux rope system is affected by inverse flux feeding and whether inverse flux feeding is also able to cause solar eruptions. In this paper, we carry out 2.5-dimensional numerical simulations to investigate the evolution of coronal flux ropes associated with inverse flux feeding in both the BPS and the HFT configurations, and then compare our simulation results with those associated with normal flux feeding in previous studies, so as to expand and improve the flux feeding mechanism of solar eruptions. The rest of this paper is arranged as follows: the numerical model used in our simulation is introduced in \sect{sec:model}, simulation results for the BPS and the HFT configurations are presented in \sect{sec:bps} and \sect{sec:hft}, respectively, and the discussion and conclusion are given in \sect{sec:dc}.

\section{Numerical model}
\label{sec:model}
\subsection{Basic equations}
\label{sec:equations}
In our 2.5-dimensional simulations, all the quantities satisfy $\partial/\partial z=0$, so that the magnetic field can be written in the following form:
\begin{align}
\textbf{B}=\triangledown\psi\times\hat{\textbf{\emph{z}}}+B_z\hat{\textbf{\emph{z}}},\label{equ:mf}
\end{align}
where $\psi$ is the magnetic flux function. With this form, the divergence-free condition of the magnetic field, $\triangledown\cdot \textbf{B}=0$, is always satisfied. The MHD equations could then be cast in the non-dimensional form:
\begin{align}
&\frac{\partial\rho}{\partial t}+\triangledown\cdot(\rho\textbf{\emph{v}})=0,\label{equ:cal-st}\\
\nonumber &\frac{\partial\textbf{\emph{v}}}{\partial t}+\frac{2}{\rho\beta_0}(\vartriangle\psi\triangledown\psi+B_z\triangledown B_z+\triangledown\psi\times\triangledown B_z)+\textbf{\emph{v}}\cdot\triangledown\textbf{\emph{v}}\\ 
&~~~+\triangledown T +\frac{T}{\rho}\triangledown\rho+g\hat{\textbf{\emph{y}}}=0,\\
&\frac{\partial\psi}{\partial t}+\textbf{\emph{v}}\cdot\triangledown\psi-\eta\vartriangle\psi=0,\\
&\frac{\partial B_z}{\partial t}+\triangledown\cdot(B_z\textbf{\emph{v}})+(\triangledown\psi\times\triangledown v_z)\cdot\hat{\textbf{\emph{z}}}-\eta\vartriangle B_z=0,\\
\nonumber &\frac{\partial T}{\partial t}-\frac{\eta(\gamma-1)}{\rho R}\left[(\vartriangle\psi)^2+|\triangledown\times(B_z\hat{\textbf{\emph{z}}})|^2 \right]\\
&~~~+\textbf{\emph{v}}\cdot\triangledown T +(\gamma-1)T\triangledown\cdot\textbf{\emph{v}}=0,\label{equ:cal-en}
\end{align}
where
\begin{align}
\vartriangle\psi=\frac{\partial^2\psi}{\partial x^2}+\frac{\partial^2\psi}{\partial y^2},~~\vartriangle B_z=\frac{\partial^2 B_z}{\partial x^2}+\frac{\partial^2 B_z}{\partial y^2}.
\end{align}
Here $\rho$ , $\textbf{\emph{v}}$, and $T$ denote the density, the velocity, and the temperature, respectively; the subscript $x, y, z$ represent the $x$, $y$, and $z-$component of the quantities; the polytropic index is $\gamma=5/3$; $g$ and $\eta$ are the normalized gravity and the resistivity, respectively; $\beta_0=2\mu_0\rho_0RT_0L_0^2/\psi_0^2=0.1$ is the characteristic ratio of the gas pressure to the magnetic pressure, where $\rho_0=3.34\times10^{-13}\mathrm{~kg~m^{-3}}$, $T_0=10^6\mathrm{~K}$, $L_0=10^7\mathrm{~m}$, and $\psi_0=3.73\times10^3\mathrm{~Wb~m^{-1}}$ are the characteristic values of of the quantities. The equations introduced above are then solved by the multi-step implicit scheme \citep{Hu1989a,Hu2003a} to simulate the evolution of the coronal magnetic system. The numerical domain is $0<x<200$ Mm, $0<y<300$ Mm; it is discretized into 400$\times$600 uniform meshes. At the left side of the domain ($x=0$), symmetric boundary condition is used. Except during the flux feeding process (which will be introduced in \sect{sec:method}), the lower boundary is always fixed; this implies that the lower boundary corresponds to the photosphere. At the other boundaries, increment equivalent extrapolation is used \citep[e.g.,][]{Zhang2020}:
\begin{align*}
U^{n+1}_{b}=U^{n+1}_{b-1}+U^{n}_{b}-U^{n}_{b-1}.
\end{align*}
Here $U$ represents the quantities in our simulation, including $\rho$, $\textit{\textbf{v}}$, $\psi$, $B_z$ and $T$; the superscript $n$ and $n+1$ indicate the quantities at the current and the next time steps, respectively; the subscript $b$ and $b-1$ indicate the quantities at the boundary, and those at the location next to the boundary, respectively. The radiation and the heat conduction in the energy equation are neglected. 

\subsection{Simulating procedures}
\label{sec:method}
The initial states in our simulations for the BPS and the HFT cases in our simulations are the same as those in \pone ~and \ptwo, respectively. The magnetic properties of a coronal magnetic flux rope could be characterized by the axial magnetic flux passing through the cross section of rope, $\Phi_z$, and the annular magnetic flux of the rope per unit length along $z$-direction, $\Phi_p$. For the BPS initial state illustrated in \fig{fig:old}(a), $\Phi_{z0}^B=9.31\times10^{19}$ Mx and $\Phi_{p0}^B=1.49\times10^{10}$ Mx cm$^{-1}$; for the HFT initial state in \fig{fig:old}(g), $\Phi_{z0}^H=4.37\times10^{19}$ Mx and $\Phi_{p0}^H=1.19\times10^{10}$ Mx cm$^{-1}$. The simulating procedures in our simulations are also similar as those in \pone ~and \ptwo: starting from the corresponding initial state, we let a small flux rope emerge from the lower base of the initial states and then interact with the pre-existing flux ropes, representing the scenario of flux feeding. In the rest of this paper, the pre-existing large flux rope is termed ``major rope'' for simplicity.
\par
The emergence of the small rope is achieved by the following procedures: the emergence begins at $t=0$ and ends at $t=\tau_E$ (for the BPS cases in \pone, $\tau_E=30\tau_A$, where $\tau_A$=17.4 s; for the HFT cases in \ptwo, $\tau_E=60\tau_A$); during this period, the small rope emerges from right below the major rope. The small rope emerges at a constant speed $v_E=2a/\tau_E$, where $a=5$ Mm is the radius of the small rope. During the emergence, e.g., at time $t_1$ ($0\leqslant t_1\leqslant \tau_E$), the emerged part of the small rope at the lower base is located within $-x_E\leqslant x\leqslant x_E$, where $x_E=(a^2-h_E^2)^{1/2}$, $h_E=a(2t_1/\tau_E-1)$. By adjusting the quantities at the lower boundary ($y=0, -x_E\leqslant x\leqslant x_E$), the emergence of the small rope is achieved:
\begin{align}
&\psi(t,x,y=0)=\psi_i(x,y=0)+\psi_E(t,x),\\
&\psi_E(t,x)=\frac{C_E}{2}\mathrm{ln}\left(\frac{2a^2}{a^2+x^2+h_E^2}\right)\label{equ:psi},\\ 
&B_z(t,x,y=0)=-C_Ea(a^2+x^2+h_E^2)^{-1}\label{equ:bz},\\
&v_y(t,x,y=0)=v_E=2a/\tau_E,~v_x(t,x,y=0)=v_z(t,x,y=0)=0,\\
&T(t,x,y=0)=2\times10^5\mathrm{~K},~\rho(t,x,y=0)=1.67\times10^{-12}\mathrm{~kg~m^{-3}}.
\end{align}
Here $\psi_i(x,y=0)$ is the magnetic flux function of the initial state at the lower boundary; $\psi_i(x,y=0)$ in \fig{fig:old}(a) and \fig{fig:old}(g) have been given in \pone ~and \ptwo, respectively. The parameter $C_E$ determines the intensity of flux feeding: the larger $C_E$, the stronger the magnetic field strength within the small rope, so that more magnetic flux is injected into the major rope, as suggested in \pone ~and \ptwo. In this paper, the given dimensionless values of $C_E$ are in the unit of $0.373\times10^{10}\mathrm{~Mx~cm^{-1}}$. Anomalous resistivity is used in the simulations in \pone ~and \ptwo:
\begin{align}
\eta=
\begin{cases}
0,& ~j\leq j_c\\
\eta_m\mu_0v_0L_0(\frac{j}{j_c}-1)^2.& ~j> j_c \label{equ:res}
\end{cases}
\end{align}
Here $L_0=10^7$ m, $v_0=128.57$ km s$^{-1}$, and $\mu_0$ is the vacuum magnetic permeability. For the BPS cases in \pone, $\eta_m=10^{-4}$ and $j_c=2.37\times10^{-4}$ A m$^{-2}$; for the HFT cases in \ptwo, $\eta_m=9.95\times10^{-2}$ and $j_c=4.45\times10^{-4}$ A m$^{-2}$.
\par
In our simulations, the values of the corresponding parameters ($a$, $\tau_E$, $\eta_m$, $j_c$, ...) for the BPS and the HFT cases are the same as those in \pone ~and \ptwo, respectively, except that the axial component of the magnetic field within the emerging small rope is negative (Eq. \ref{equ:bz}), so that the direction of $B_z$ in the small rope is opposite to the major rope. In this way, we may explicitly compare the influence of the inverse flux feeding processes on coronal flux rope systems with that of the normal ones investigated in \pone ~and \ptwo.

\section{Simulation results}
\label{sec:result}

\subsection{BPS configuration}
\label{sec:bps}
\begin{figure*}
	\includegraphics[width=\hsize]{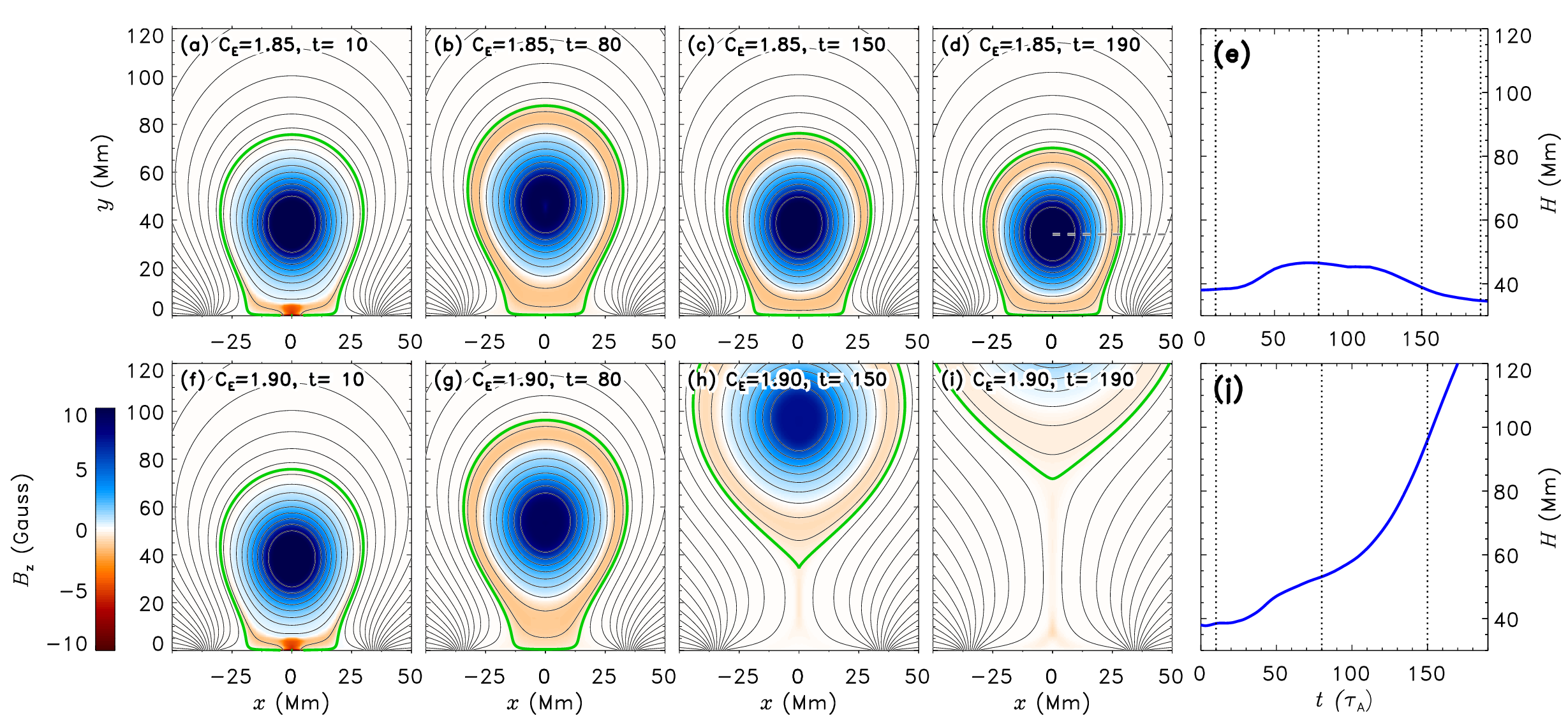}
	\caption{Simulation results of the flux rope in the BPS configuration. Panels (a)-(d) illustrate the evolution of the magnetic configuration for the case with $C_E=$1.85, and panel (e) plot the evolution of the height of the rope axis, with the vertical dotted lines corresponding to the times of Panels (a)-(d). Panels (f)-(l) are the results for the case with $C_E=$1.90. The meanings of the symbols and colors are the same as those in \fig{fig:old}.}\label{fig:bps}
\end{figure*}
The simulation results of typical inverse flux feeding processes in the BPS configuration are illustrated in \fig{fig:bps}; the top row and the bottom row show the cases with $C_E=$1.85 and 1.90, respectively. After the onset of the inverse flux feeding process, the emerging small rope interacts and merge with the lower section of the major flux rope, which is similar as the process in, e.g., \cite{Linton2001a} and \cite{Linton2006}; as a result, the negative axial magnetic flux within the small rope is injected into the major rope, as shown by the red color in \fig{fig:bps}(a) and \ref{fig:bps}(f). The injected flux is then transported across the major rope (\fig{fig:bps}(a)-\ref{fig:bps}(b) and \fig{fig:bps}(f)-\ref{fig:bps}(g)), and the magnetic configuration of the resultant major rope after inverse flux feeding (\fig{fig:bps}(b) and \fig{fig:bps}(g)) is interesting: the injected negative axial flux does not completely cancel out with the pre-existing positive axial flux within central region of the major rope, but is eventually dispersed only within the outer section of the major rope, resulting in a double-layer configuration. It is noteworthy that the positive axial magnetic flux is concentrated in the central region of the major rope in the initial state (\fig{fig:old}(a)). Since the negative axial flux injected by inverse flux feeding is primarily distributed in the outer section of the resultant major rope, the cancellation between the oppositely directed axial fluxes should be limited. The spatial separation of the pre-existing positive and the injected negative axial fluxes leads to the formation of the double-layer configuration.

\par
The subsequent evolutions of the major flux rope after flux feeding in the cases with $C_E=$1.85 and 1.90 are quite different. The major rope remains sticking to the photosphere in the case with $C_E=$1.85 (\fig{fig:bps}(c)-\ref{fig:bps}(d) and \fig{fig:bps}(e)), indicating that this is a non-eruptive case. In the case with $C_E=$ 1.90, however, the major rope keeps rising after flux feeding, resulting in a full eruption of the major rope (\fig{fig:bps}(h)-\ref{fig:bps}(i) and \fig{fig:bps}(j)). The total axial current in the major rope during the onset of the eruption is of the order $7\times10^{10}$ A, which is close to that during the eruption caused by normal flux feeding (\fig{fig:old}(a)-\ref{fig:old}(f)), indicating similar strapping field strength. It is interesting that the major rope remains in the double-layer configuration during its whole evolution, with the positive and negative axial flux separated from each other. 
\par
\begin{figure*}
	\includegraphics[width=0.9\hsize]{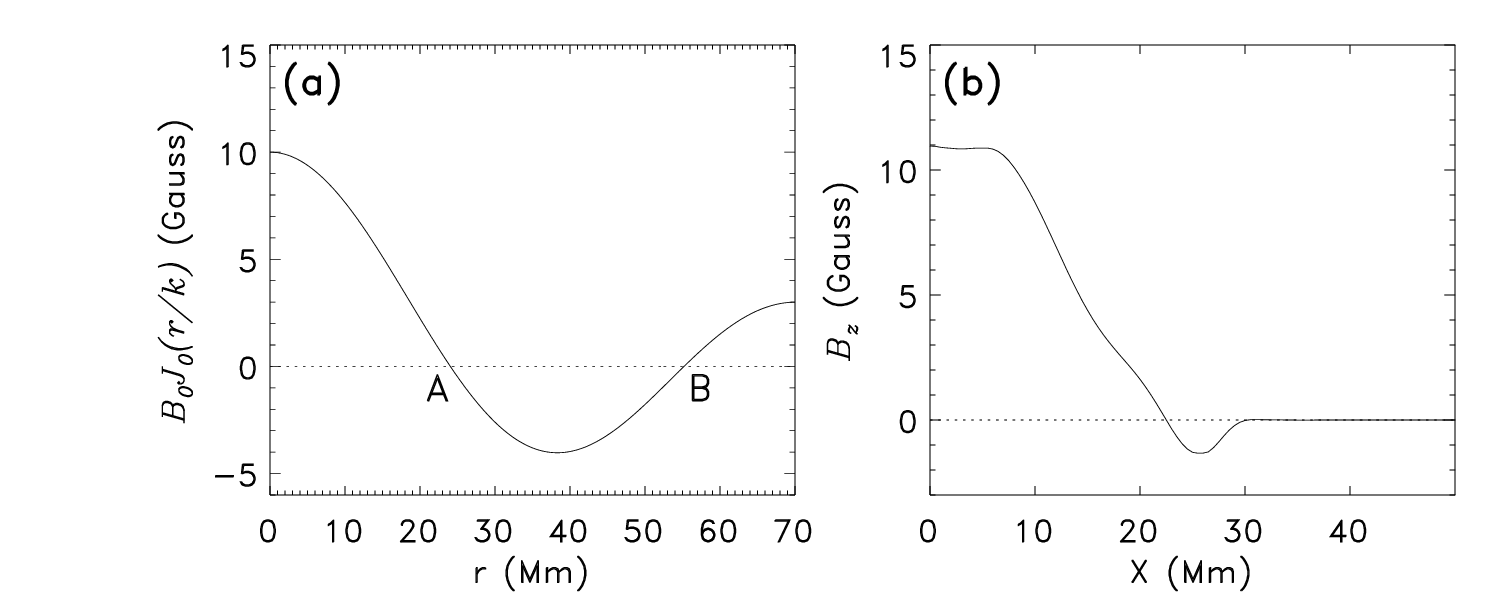}
	\caption{Comparison of the radial distribution of $B_z$ predicted by model and that in our simulation results. Panel (a) is the distribution of $B_z$ predicted by Lundquist solution, in which ``A'' and ``B'' mark the zero point; panel (b) is the distribution of $B_z$ along the gray dashed line in \fig{fig:bps}(d).}\label{fig:bz}
\end{figure*}

The resultant double-layer configuration of the major rope after inverse flux feeding might be explained by force-free flux rope model. For one-dimensional force-free flux tubes, \cite{Lundquist1951} gave a flux rope model with Lundquist solution:
\begin{align}
B_\phi(r)=B_0J_1(\frac{1}{k}r),~~B_z(r)=B_0J_0(\frac{1}{k}r),
\end{align}
where $r$ is the radial distance from the rope axial. An example of the radial distribution of the axial component of the magnetic field, $B_z(r)$, predicted by Lundquist solution is plotted in \fig{fig:bz}(a) (assuming $B_0=$10 G and $k=100$ Mm). Obviously, the axial component of the magnetic field reverses direction at the zeros of $B_z(r)$, which is usually regarded as undesirable feature for solar applications. In our simulations, however, we find this kind of the $B_z$ profile could exist within the flux rope:  \fig{fig:bz}(b) plots the distribution of $B_z$ along the gray dashed line in \fig{fig:bps}(d); the $B_z$ profile within the rope in our simulation is similar as that within the second zero point (marked by ``B'' in \fig{fig:bz}(a)) of $B_z(r)$ in Lundquist solution. Therefore, Lundquist solution might explains the double-layer equilibrium state in our simulation, and our simulation results in turn indicate that the axial magnetic field reversals could exist within solar magnetic flux ropes. We note that our simulation is not force-free, and the flux rope in our simulation is not one-dimensional, so that the distribution of $B_z$ in our simulation does not exactly follow that predicted by Lundquist solution. 

\par
To investigate influence of inverse flux feeding on coronal flux ropes, we calculate the magnetic fluxes of the resultant major flux rope at t=30$\tau_A$, and compare them with those of the initial BPS state, i.e., $\Phi_{z0}^B$ and $\Phi_{p0}^B$. It is demonstrated that the poloidal magnetic flux $\Phi_{p}$ remains unchanged after inverse flux feeding in both the case with $C_E=$1.85 and 1.90, whereas the axial magnetic flux $\Phi_z$ decreases to 6.68$\times10^{19}$ Mx and 6.43$\times10^{19}$ Mx in the case with $C_E=$1.85 and 1.90, respectively. We note that what we calculate above is the total signed axial magnetic flux of rope. Since there are both positive and negative axial fluxes distributed within the resultant major rope (\fig{fig:bps}(b) and \fig{fig:bps}(g)), we also calculate the total unsigned axial magnetic flux, $|\Phi_z|$. For the initial BPS state, $|\Phi_z|_0^B=\Phi_{z0}^B=9.31\times10^{19}$ Mx; for the resultant rope after inverse flux feeding, $|\Phi_z|$ increases to 1.15$\times10^{20}$ Mx and 1.17$\times10^{20}$ Mx in the case with $C_E=$1.85 and 1.90, respectively. Obviously, the total unsigned axial flux accumulates after inverse flux feeding. 
\begin{figure*}
	\includegraphics[width=\hsize]{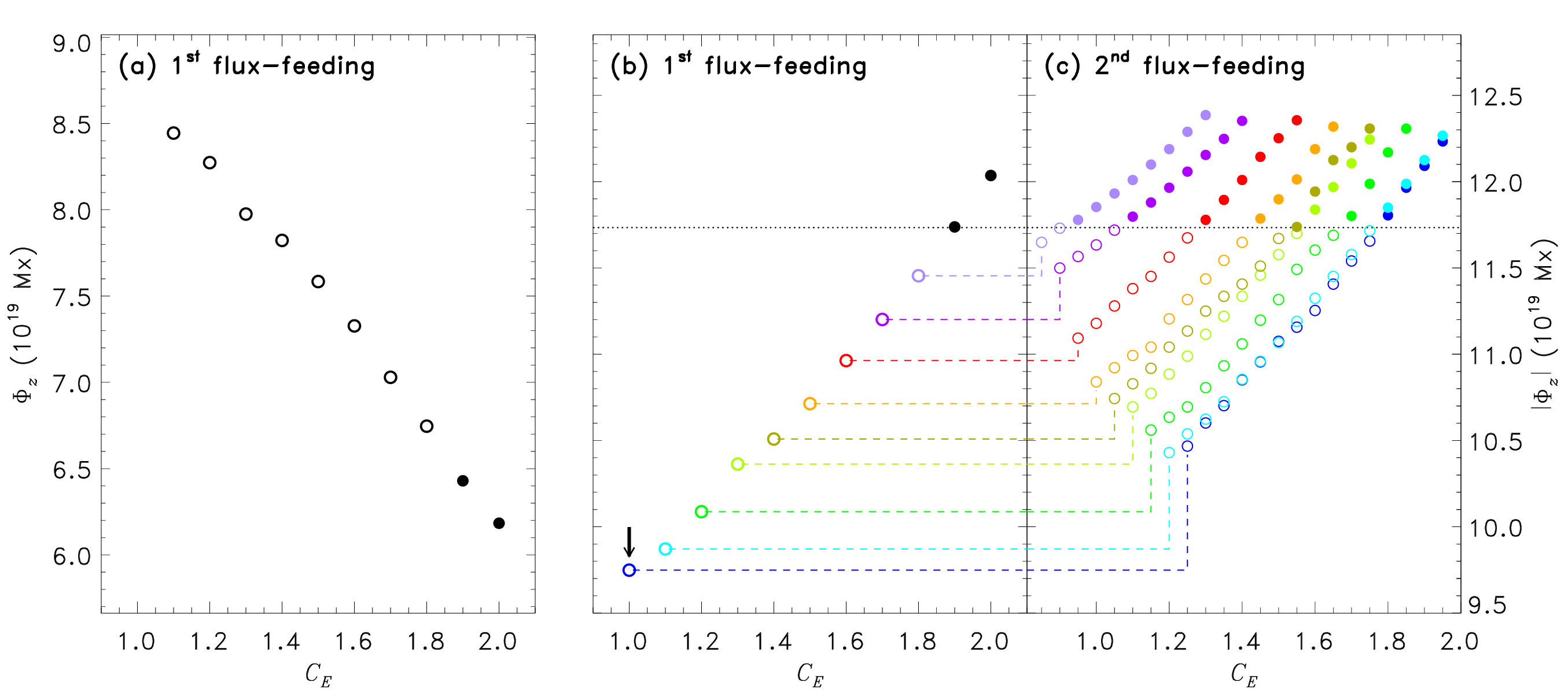}
	\caption{Axial magnetic fluxes of the resultant major rope after inverse flux feeding with different $C_E$. $\Phi_z$ and $|\Phi_z|$ of the resultant ropes after the 1$st$ round of inverse flux feeding are plotted in panel (a) and (b), respectively, and panel (c) are $|\Phi_z|$ of the resultant ropes after the 2$nd$ round of inverse flux feeding. The eruptive cases are plotted in dots, and the non-eruptive cases in small circles. The correspondence between the cases in panel (c) and the corresponding new initial state is indicated by the colored dashed line.}\label{fig:bpscri}
\end{figure*}
\par
Following \pone, we simulate many cases with different $C_E$, so as to investigate what initiates the eruption. The signed axial fluxes of the corresponding resultant major rope are plotted in \fig{fig:bpscri}(a), and the unsigned axial fluxes in \fig{fig:bpscri}(b). For the case with larger $C_E$, i.e. stronger intensity of inverse flux feeding, the total signed axial flux $\Phi_z$ of the resultant rope is smaller, whereas the unsigned aixal flux $|\Phi_z|$ is larger, indicating that more negative flux is injected into the major rope. Moreover, $|\Phi_z|$ of the resultant rope in the eruptive cases (dots in \fig{fig:bpscri}(b)) tends to be larger than that in the non-eruptive cases (small circles in \fig{fig:bpscri}(b)). Thus we may infer that the total unsigned axial flux $|\Phi_z|$ rather than the signed flux $\Phi_z$ should play a decisive role in triggering the eruption of the flux rope. To confirm this, we further simulate many other cases, the initial states of which are changed to the resultant flux ropes in the non-eruptive cases plotted in \fig{fig:bpscri}(b). For simplicity, we call these cases as ``second round of inverse flux feeding'', which are plotted in \fig{fig:bpscri}(c), and those plotted in \fig{fig:bpscri}(b) are called ``the first round of inverse flux feeding''. The second round of flux feeding cases and their corresponding initial states are plotted by the same color, and are also connected by the colored dashed lines. Combining \fig{fig:bpscri}(b) and \fig{fig:bpscri}(c), it is suggested that the eruptive (dots) and the no-eruptive (small circles) cases are well separated, i.e., there should be a critical value of the unsigned axial flux $|\Phi_z|_c^B$ of the order 1.17$\times10^{20}$ Mx, as marked by the horizontal dotted line in \fig{fig:bpscri}(b)-\ref{fig:bpscri}(c). This value is very close to the critical $\Phi_z$ associated with normal flux feeding found in \pone; since there is only positive magnetic flux within the resultant rope after normal flux feeding (\fig{fig:old}), $|\Phi_z|$ always equals to $\Phi_z$ in the simulation results in \pone, indicating that the critical $|\Phi_z|$ associated with normal (\pone) and inverse (this paper) flux feeding are very close. Therefore, we may conclude that both normal and inverse flux feeding processes are able to cause coronal flux rope eruptions in BPS configurations; what dominates the onset of the eruption is not the total singed axial flux, but the total unsigned axial flux $|\Phi_z|$: the flux rope will only erupt if its $|\Phi_z|$ surpasses the critical value $|\Phi_z|_c^B$.

\subsection{HFT configuration}
\label{sec:hft}

\begin{figure*}
	\includegraphics[width=\hsize]{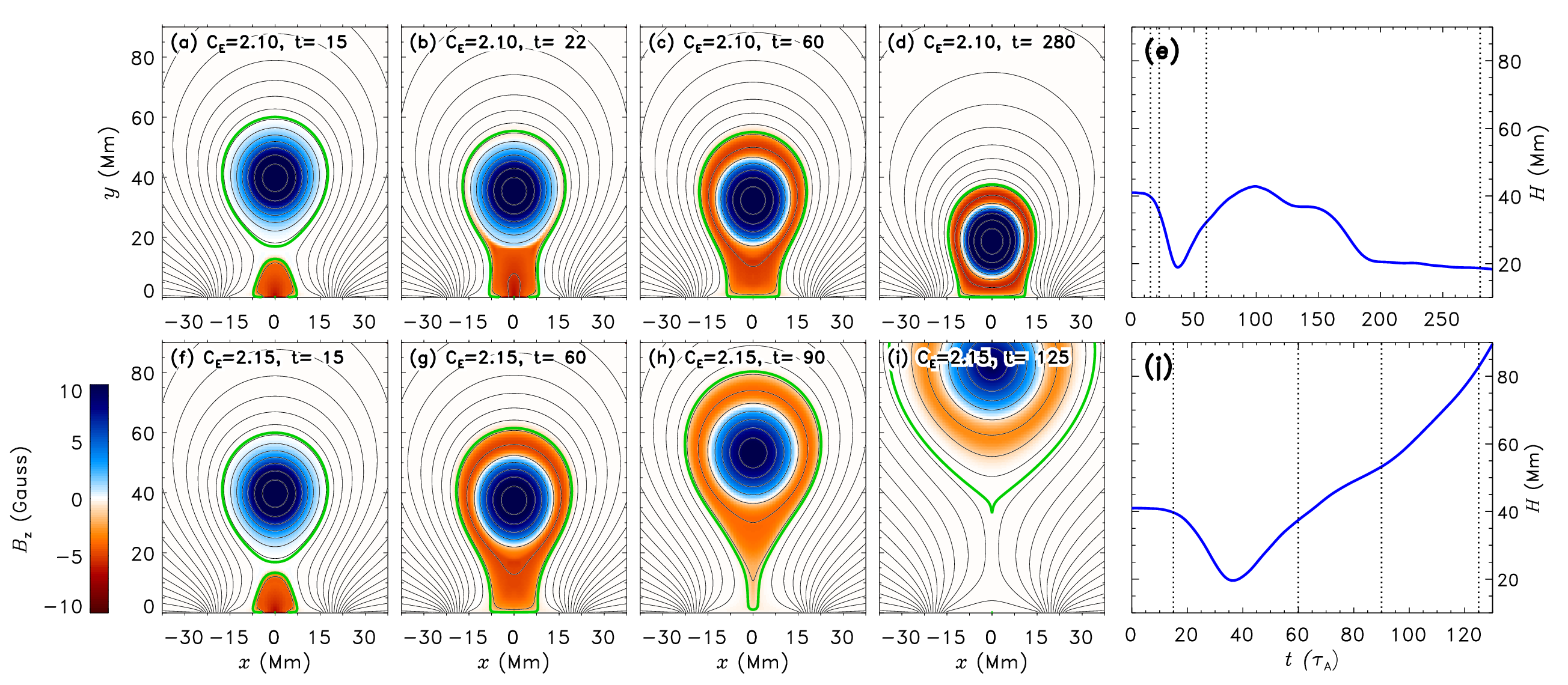}
	\caption{Simulation results of the flux rope in the HFT configuration. Top and bottom panels are the results for the case with $C_E=$2.10 and $C_E=$2.15, respectively. The meanings of the symbols and colors are the same as those in \fig{fig:old}.}\label{fig:hft}
\end{figure*}
Figure \ref{fig:hft} illustrates the simulation results of typical inverse flux feeding processes in the HFT configuration; the top row and the bottom row show the cases with $C_E=$2.10 and 2.15, respectively. After the inverse flux feeding process begins, the small rope emerges from the photosphere (\fig{fig:hft}(a)), pushing the arcades (as marked by the green curves below the major rope in \fig{fig:old}(g)) upward, which reconnect with the magnetic field of the major rope. The interaction and reconnection between the major rope and the arcades have been demonstrated in detail in \ptwo. After all the arcades reconnects with the major rope, the small emerging rope itself interacts and merges with the major rope (\fig{fig:hft}(b)), during which negative axial magnetic flux carried by the small rope is injected into the major rope. As illustrated in \fig{fig:hft}(c)-\ref{fig:hft}(d), the injected flux is eventually distributed within the outer section of the major rope, so that results in a double-layer configuration, which is similar as that in the BPS cases (e.g. \fig{fig:bps}(c)). It is demonstrated that the case with $C_E=$2.10 is non-eruptive: the flux rope eventually falls down to the photosphere (\fig{fig:hft}(d)), so that the HFT configuration collapses. In the case with $C_E=$2.15, however, the resultant rope keeps rising after inverse flux feeding, so that this is an eruptive case. Following \ptwo, we also calculate magnetic fluxes of the resultant major rope: for the case with $C_E=2.10$, $\Phi_z=5.78\times10^{18}$ Mx, $|\Phi_z|=7.05\times10^{19}$ Mx, $\Phi_p=1.12\times10^{10}$ Mx cm$^{-1}$; for the case with $C_E$=2.15, $\Phi_z=7.32\times10^{17}$ Mx, $|\Phi_z|=7.58\times10^{19}$ Mx, $\Phi_p=1.12\times10^{10}$ Mx cm$^{-1}$. Obviously, $\Phi_z$ decreases but $|\Phi_z|$ accumulates after inverse flux feeding; since larger $C_E$ implies stronger intensity of inverse flux feeding, more negative axial flux is injected, resulting in larger $|\Phi_z|$ of the resultant rope. In both of these two cases, the poloidal flux of the major rope is reduced by about $\bigtriangleup\Phi_p=0.07\times10^{10}$ Mx cm$^{-1}$ after inverse flux feeding. As discussed in \ptwo, the reduced poloidal flux should be caused by the reconnection between the arcades below the rope (\fig{fig:old}(g)) and the major rope, which peels off the outermost section of the major rope. The total axial current in the major rope during the onset of the eruption in the case with $C_E=$2.15 is of the order  $7\times10^{10}$ A, which is also close to that during the eruption caused by normal flux feeding (\fig{fig:old}(h)-\ref{fig:old}(l)).
\par
We also further simulate many cases with different $C_E$, and discover that the eruptive and non-eruptive cases are also well separated: the case with $C_E\leq$ 2.10 (for the corresponding resultant major rope, $|\Phi_z|\leq7.05\times10^{19}$ Mx) is non-eruptive, whereas $C_E\geq$ 2.11 (for the corresponding resultant major rope, $|\Phi_z|\geq7.19\times10^{19}$ Mx) is eruptive. This indicates that there is a critical unsigned axial flux of the order $=7.10\times10^{19}$ Mx, which is almost the same as that found in \ptwo. Therefore, both normal and inverse flux feeding processes are able to cause the eruption of coronal flux ropes in the HFT configuration, provided that the critical $|\Phi_z|$ is reached after flux feeding. This conclusion is quite similar as that for the BPS cases.

\section{Discussion and conclusion}
\label{sec:dc}
In this paper, we investigate the influence of inverse flux feeding on coronal magnetic flux rope systems. During inverse flux feeding processes, newly emerging magnetic flux directly interact with pre-existing coronal magnetic flux rope. As a result, axial magnetic flux, whose direction is opposite to that within the pre-existing major flux rope, is injected into the rope. The injected axial flux is distributed within the outer section of the major rope (\fig{fig:bps} and \fig{fig:hft}), so that the total signed axial flux of the major rope decreases but the total unsigned axial flux increases after inverse flux feeding. Our simulation results suggest that the onset of the eruption should be associated with the total unsigned axial flux of the major rope: if the amount of the axial flux injected by inverse flux feeding is large enough so that the unsigned axial flux of the major rope exceeds a critical value, the eruption of the coronal flux rope is initiated. As discussed in \sect{sec:bps} and \sect{sec:hft}, although the signed axial fluxes after normal and inverse flux feeding are quite different, the values of the critical unsigned axial flux for inverse flux feeding are very close to those for normal flux feeding in both the BPS and the HFT cases. This indicates that it is not the signed but the unsigned axial flux that dominates the onset of the eruption, and the critical unsigned axial flux is almost the same regardless of whether normal or inverse flux feeding occurs in the flux rope system. Therefore, we conclude that both normal and inverse flux feeding are efficient in causing coronal flux rope eruptions, provided that the critical unsigned axial flux of the rope is reached.
\par
\begin{figure*}
	\includegraphics[width=\hsize]{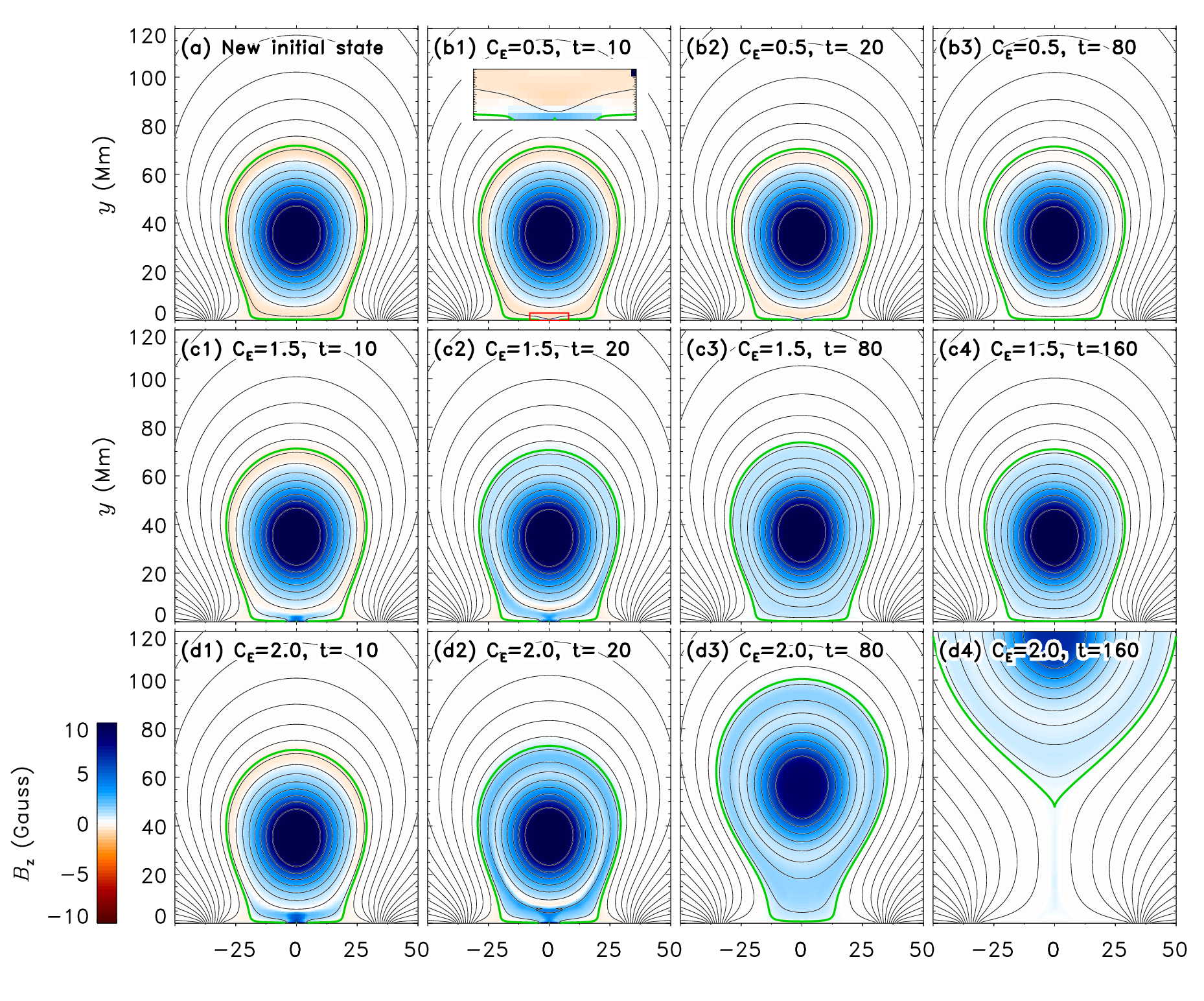}
	\caption{Simulation results about the flux feeding processes in a flux rope system with double-layer configuration. Panel (a) is the new initial state. Panels (b1)-(b3) are the evolution for the case with $C_E=$0.5, and the inset in panel (b1) correspond to the region marked by the red box in panel (b). Panels (c1)-(c4) and panels (d1)-(d4) are the results for the case with $C_E=$1.5 and $C_E=$2.0, respectively. The meanings of the symbols and colors are the same as those in \fig{fig:old}.}\label{fig:add}
\end{figure*}
To further investigate the influence of flux feeding on the unsigned axial flux of coronal flux ropes, we simulate several additional cases, as shown in \fig{fig:add}. Here we switch to use a new initial state, which is the resultant equilibrium state after the first round of inverse flux feeding in the BPS configuration with $C_E=1.0$ (corresponding to the small blue circle marked by the small arrow in \fig{fig:bpscri}(b)). \fig{fig:add}(a) illustrates this new initial state, in which $|\Phi_z|=9.75\times10^{19}$ Mx. Obviously, there is negative axial flux distributed within the outer section of the major rope, which is different from the initial state of the simulation in \sect{sec:bps} (\fig{fig:old}(a)). Starting from this new initial state, we let a small flux rope containing positive axial flux emerge from the lower boundary, which is simply achieved by reversing the minus sign in Eq. \ref{equ:bz}. Figures \ref{fig:add}(b1)-\ref{fig:add}(b3) illustrate the corresponding simulation result with $C_E=0.5$: positive axial flux is injected into the major rope (\fig{fig:add}(b1) and the inset), and cancels out with the pre-existing negative axial flux in the outer section of the major rope (\fig{fig:add}(b2)-\ref{fig:add}(b3)). As a result, $|\Phi_z|$ decreases to $9.29\times10^{19}$ Mx after this flux feeding process, and the rope does not erupt after flux feeding; in fact, the resultant rope in this case should be even further from the onset of the eruption than its initial state in \fig{fig:add}(a). This indicates that flux cancellation is possible during flux feeding when axial flux is present near the boundary of the flux rope, leading to a decrease rather than an accumulation of the total unsigned axial flux after flux feeding. It is noteworthy that this research focuses on the influence of flux feeding on the total magnetic flux of the flux rope, rather than on the detailed magnetic reconnection process of the oppositely directed axial flux, which could hardly be investigated under the translational invariance assumption in 2.5-Dimensional simulations. For stronger flux feeding process with $C_E=1.5$, not only the pre-existing negative axial flux is cancelled out with the injected positive flux (\fig{fig:add}(c1)), but also extra positive axial flux is injected into and then distributed within the outer section of the major rope (\fig{fig:add}(c2)-\ref{fig:add}(c3)). For this case with $C_E=1.5$, $|\Phi_z|$ increases to $1.15\times10^{20}$ Mx after flux feeding, but is still smaller than the critical unsigned axial flux $|\Phi_z|_c^B\sim1.17\times10^{20}$ Mx, so that the major rope does not erupt (\fig{fig:add}(c4)). For even stronger flux feeding process with $C_E=2.0$ (\fig{fig:add}(d1)-\ref{fig:add}(d4)), $|\Phi_z|$ of the resultant rope is $1.22\times10^{20}$ Mx, larger than $|\Phi_z|_c^B$, so that the major rope eventually erupts. These simulation results further confirm that the unsigned axial flux is very important, but it does not always accumulates after flux feeding: both the properties of the flux feeding processes and the magnetic configuration of the pre-existing coronal flux rope should influence the unsigned axial flux of the resultant rope after flux feeding. 
\par
An interesting phenomenon found in our simulation results is that there could be both positive and negative axial magnetic field distributed within a flux rope, as shown in \fig{fig:bps} and \fig{fig:hft}, and the opposite axial magnetic field components within the flux rope also results in the coexistence of magnetic helicity with opposite signs within the rope. This kind of double-layer configuration within the flux rope is the fundamental cause of the discrepancy between the signed and unsigned axial flux of the rope. Based on our simulation results, it could be inferred that there might probably be reversals of the axial magnetic field component within the flux rope in the solar corona. This implies that a flux rope might contain magnetic helicity of opposite sign to that of the surrounding active region, and as a result, the magnetic helicity in the active region may even increase rather than decrease after the eruption of the flux rope. In fact, many previous studies have suggested that physical parameters that are related to magnetic helicity should play a critical role in the initiation of solar eruption \citep[e.g.,][]{Pariat2017,Zuccarello2018,Thalmann2019,Thalmann2020,Gupta2021}. Therefore, in our future work, we plan to build upon the present study to explore helicity-related parameters as potential thresholds for the eruption of flux rope with complex internal structures. Moreover, interplanetary magnetic flux ropes might also exhibit this kind of double-layer configuration, potentially introducing new challenges for magnetic cloud modelling \citep[e.g.,][]{Zhao2017}. More observational and theoretical studies are still needed to further investigate the detailed magnetic topology inside coronal magnetic flux ropes and magnetic clouds, and the influence of the internal topological characteristics on the evolution of coronal flux ropes and magnetic clouds.



\begin{acknowledgements}
We appreciate Dr. Xiaolei Lei for his suggestion and advice. We also appreciate the anonymous referee for the valuable comments. This research is supported by the Strategic Priority Research Program of the Chinese Academy of Sciences (Grant No. XDB0560000), the National Natural Science Foundation of China (NSFC 42188101, 42174213, 11925302, 41804161), the Informatization Plan of Chinese Academy of Sciences (CAS-WX2022SF-0103), USTC Research Funds of the Double First-Class Initiative (YD2080002011), the open subject of Key Laboratory of Geospace Environment (GE2018-01), and the Innovation Program for Quantum Science and Technology (2021ZD0300302). We also acknowledge for the support from National Space Science Data Center, National Science \verb"&" Technology Infrastructure of China (www.nssdc.ac.cn). Quanhao Zhang acknowledge for the support from Young Elite Scientist Sponsorship Program by the China Association for Science and Technology (CAST).
\end{acknowledgements}


\end{document}